# LOW POWER SI CLASS E POWER AMPLIFIER AND RF SWITCH FOR HEALTH CARE

Wei Cai[1], Jian Xu[2], and Liang Huang[3]

[1]Department of Electrical Engineering and Computer Science, University of California, Irvine, CA, USA
*caiw2@uci.edu*

[2]Division of Electrical and Computer Engineering, School of Electrical Engineering and Computer Science, Louisiana State University, Baton Rouge, LA, USA
*j.xu@aya.yale.edu*

[3]Department of Information & Electronic Engineering, ZheJiang Gongshang University, Hang Zhou, Zhejiang, China
*huangliang@zjgsu.edu.cn*

## ABSTRACT

*This research was to design a 2.4 GHz class E Power Amplifier (PA) for health care, with 0.18um Semiconductor Manufacturing International Corporation CMOS technology by using Cadence software. And also RF switch was designed at cadence software with power Jazz 180nm SOI process. The ultimate goal for such application is to reach high performance and low cost, and between high performance and low power consumption design. This paper introduces the design of a 2.4GHz class E power amplifier and RF switch design. PA consists of cascade stage with negative capacitance. This power amplifier can transmit 16dBm output power to a 50Ω load. The performance of the power amplifier and switch meet the specification requirements of the desired.*

## KEYWORDS

*Cascode, Negative Capacitance, Class E, Power amplifier, Healthcare, RF switch*

## 1. INTRODUCTION

Wireless Sensor Networks (WSN) can be widely applied to solve a vast array of problems, under varied conditions[1]. Such WSNs can provide near-real time, non-stop data over a large sampling area or population, by large distributing many devices to monitor the surrounding environment [2][3][4][5][6][7][8][9]. WSNs could deliver considerable efficiencies to otherwise with costly tasks [10][11][12]. For example, patient monitoring carries considerable cost, especially if used to a large section of the patient. Desirable solutions can be recognized leveraging WSNs and the present cellular communication. Academic and industry research is currently ongoing investigating such frameworks [13][14][15][16][17][18].

Due to current hardware components restrictions, healthcare application of WSNs are still in the early stages[19][20][21][22]. Such devices require Food and Drug Administration (FDA) approval, which can be challenging and costly due to the requirement that the devices pass a number of safety tests. Not so many companies and research institutions can successfully building such a heath care device under full FDA approval [23].

WSNs consist of a number of networked elements, which are individually called sensor nodes. Sensor nodes usually contain all kinds of hardware elements, such as batteries, sensors, antennas, memory, ADC, FPGA, etc [24][25][26][30]. A major design challenge for medical applications is that - designs a cost effective device which would meet functional requirements [2][27][28][29]. In order to implement such networks with a massive amount of nodes, each node must be low cost. Typically, each device must provide long working cycles without battery recharging. This pushes most sensor node designs to be super ultra-low power. Achieving this low power performance at low cost are critical to making such sensor networks feasible [31][32].

The main design challenge for such WSN is the high power consumption of portable devices. One possible answer to this task is the integration of the digital, analog and RF circuitry into one chip. Thus, switch and power amplifiers have different and unique characteristics, which requires different processes to tape out each one. But switches and PAs can be integrated on a single SIP. A system in package (SiP) is a module that contains multiple integrated circuits. A SiP has the same function as an electronics system. They are often used in the smart phones and PCs. Dies could be connected via bondwires between packages. Alternatively, solder bumps may be utilized in a stacked architecture in the package.

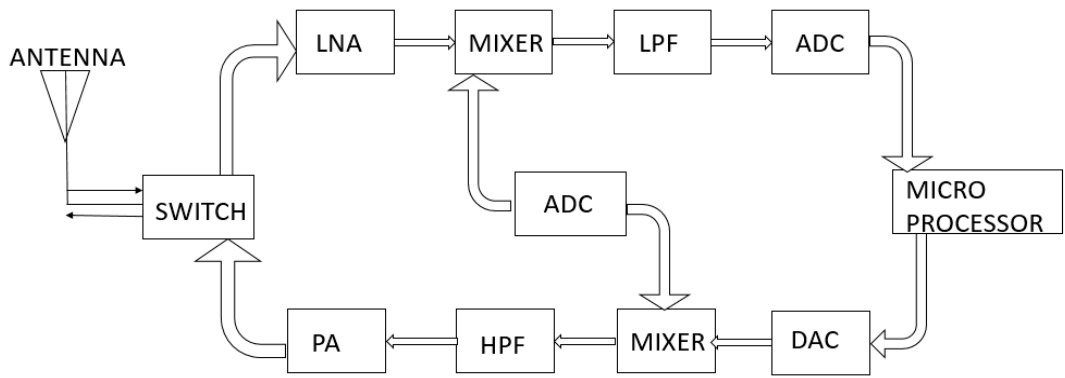

Figure 2. Block diagram of a transmitter

The receiver will receive the signal and will also perform DSP processing after the data is sent out by the transmitter [32]. Figure 2 is the basic transmitter block. It is always desirable that the transmitter and receiver are low power consumption. In order to meet the standards, the PA and RF switch are designed as shown in table 1.

Table 1: PA design requirement.

| Parameter | Target(Unit) |
| --- | --- |
| Output Power | 15 dBm |
| Power gain | 50 Db |
| Stability | >1 |
| S11 | -10 dB |
| Insertion | 0-1.2 Db |
| Isolation | >40 Db |
| IIP3 | 55 Db |

## 2. METHODS

CMOS radio-frequency (RF) front-end circuits has developed extremely fast over the past 30 years. Getting the trade-offs between high performance and low cost, and between high performance and low power consumption design, people always try to achieve these target [34].

The cascode circuit with negative capacitance is shown in the figure 2. The advantage of this structure is that it provides less parasitic capacitance, since it allows the parasitic capacitance to be tuned at the driver stage. A shunt inductor instead of capacitor can also be inserted at the driver stage to filter out the unwanted parasitic capacitance, at the cost of wafer area [33].

As seen in figure 2, a negative capacitance can be implemented by a capacitor with a common gate amplifier. For class E power amplifiers, transistor M1 acts as a switch. Transistor M2 delivers high gain, when biased at saturation [35].

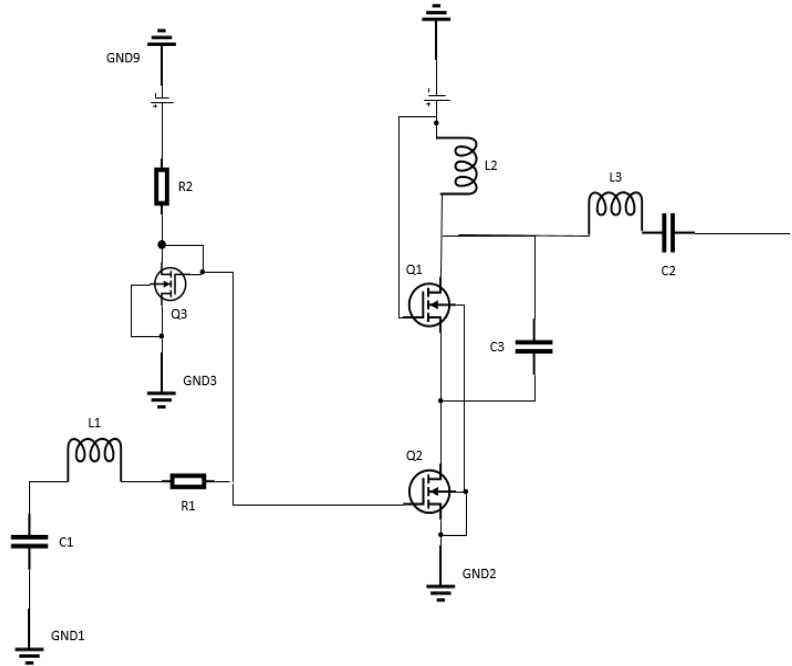

Figure 2. Block diagram of a class E power amplifier

To get the optimum bias, cadence simulation such as PSS are completed. Detailed design values can be seen in Table 2.

Table 2: 2.4GHz PA driver stage component.

| Parameter | Size (Unit) |
| --- | --- |
| Q1 | W/L=0.3um/0.6um (f=66,m=24) |
| Q2 | W/L=0.3um/0.6um (f=66,m=24) |
| Q3 | W/L=0.8um/0.6um (f=4,m=2) |
| L1 | 36nH (Q=20) |
| L2 | 20nH (Q=20) |
| L3 | 20nH (Q=20) |
| C1 | 240fH |
| C2 | 600fH |
| C3 | 11pF |
| R1 | 10.5ohm |
| R2 | 3.8Kohm |

FET switches usually have three different topologies, such as series, shunt and combinatorial topology. Due to modern, complicated requirements, users for in health care usually require the combinatorial topology to meet their stringent requirements, as seen in the Figure 3. When a control voltage is set high, the series FET would be on, which means a signal would pass to the following transistor, where a shunt FET would connect to the ground. When the control is set low, the series FET is off, so there will be no signal flow through the transistor, however the shunt FET will pass the signal.

In terms of RF switch performance, there are several key parameters, such as reflection coefficient S11, insertion loss S21 and isolation S31[34][35]. S11 is the input reflection coefficient, which is voltage ratio of the reflected wave on the input port to the original wave. This parameter represents the power loss from impedance mismatches, also known as voltage standing wave ratio (VSWR). S21 represents the forward voltage gain. A low insertion loss between source and active switch is critical to increase the efficiency. S31 is also a very important switch parameter. When there is 3 ports, two ports are on, and another port is off, and this parameter is a measure of the transmission coefficient from the source to the off arm. This parameter represents how much power was leaked into the off arm. Besides the S11, S21, S31, a switch design's value must also consider the intercept point (IP3). This parameter is a measure of the linearity of a device, which also known as intermediation distortion. The third order intercept point is where the intercept of the fundamental frequency and the third order of the fundamental frequency. There are many other parameters that can be considered, but the S-parameters and the IP3 are the critical ones that must often be considered. A good RF switch usually possesses: low insertion loss, high isolation, high power handling, and very high ESD immunity. A shunt FET would connect to the ground. When the control is set low, the series FET is off, so there will be no signal flow through the transistor, however the shunt FET will pass the signal.

FETs are three terminal devices are usually fabricated as SOI or GaAs. The basic function is shown in Fig. 3[34]. When the gate is more positively biased then the source, a channel would be formed between the source and drain side, so the resistance is lowered, and a current can be flowed between source and drain terminals. However, if the gate voltage is equal or smaller than the source terminal, no channel will form, resistance will be much higher, and no current flows through channel.

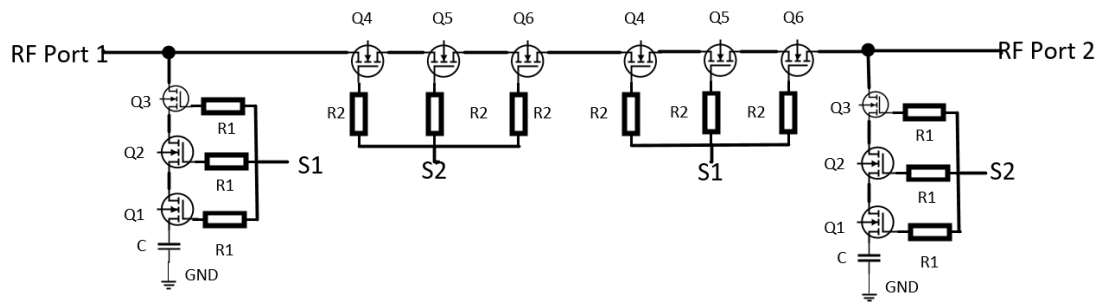

Figure 3. Modified series-shunt FET switch

## 3. RESULTS

As seen in figure 4(a), the output power was 17 dBm. As seen in figure 4(b), the S11 is less than -10 dB at 2.4 GHz frequency, also, the total power consumption is 2.061 W.

As seen in figure 5(a), Kf is larger than 1 for all the simulated frequencies, so this design is completely stable. And the power gain could reach 94 dB.

As seen in figure 6(a), at 5 GHz, an insertion loss is 1.36 dB and an isolation is 58.5 dB.

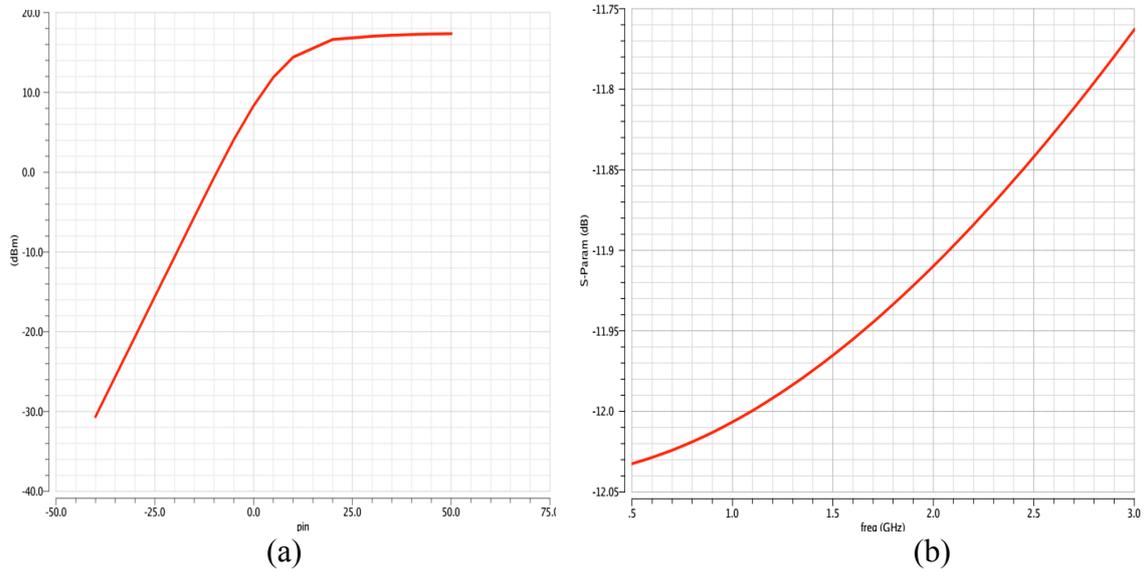

Figure 4. (a) Output power  (b) S11

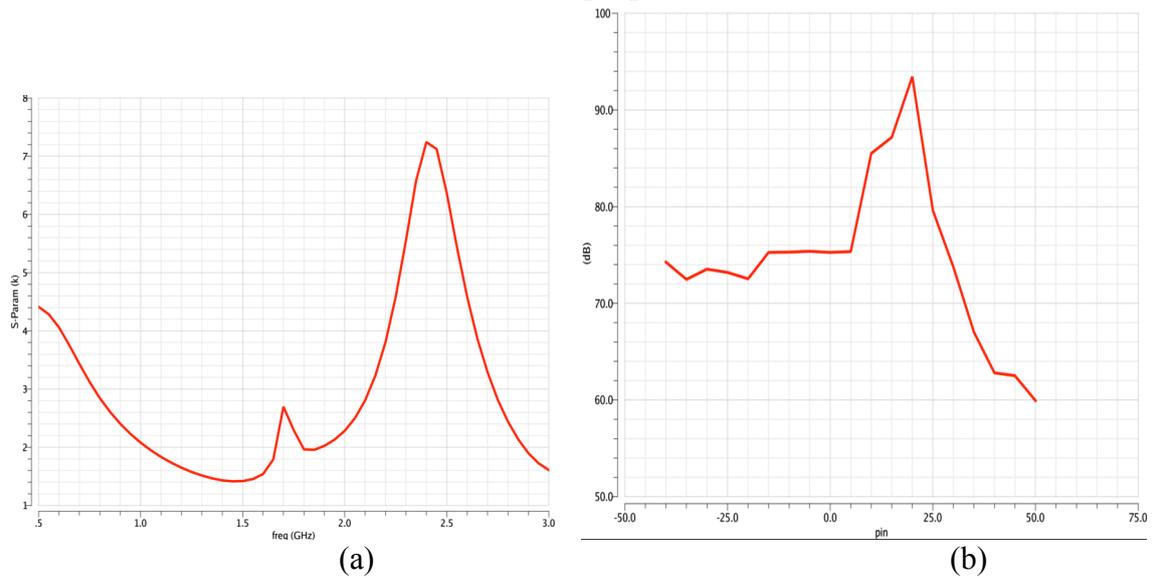

Figure 5. (a) Kf  (b)  Power gain

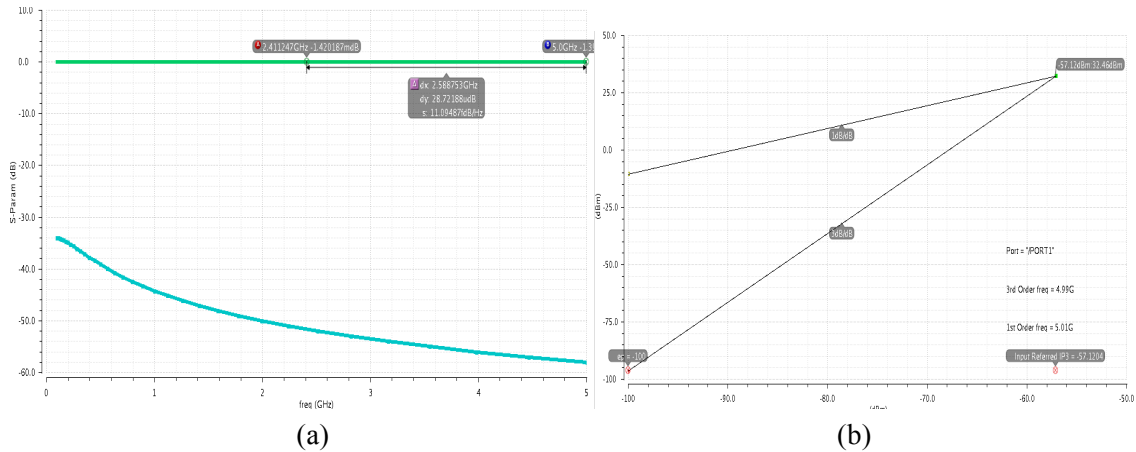

(a)  (b)

Figure 6. (a) Insertion (green) and isolation (blue**)**  (b)  IIP3

## 4. CONCLUSION

This paper describes the method of designing and simulating power amplifier in SMIC CMOS 180 nm process and RF switch at SOI process 180nm technology. This PA and switch are used for sensor networks which can be integrated at SIP. Such kind research is still under developing, to realize this a low cost and low power device, future improvements are needed.

## REFERENCES

[1]     Stults BM., (1984) "Preventive Health Care for the Elderly", Western Journal of Medicine, pp 832- 845.
[2]     Afsaneh Minaie, Ali Sanati-Mehrizy, Paymon Sanati-Mehrizy & Reza Sanati-Mehrizy (2013). "Application of Wireless Sensor Networks in Health Care System", ASEE, vol 3, pp 21-24.
[3]     Ann K Nowinski, Fang Sun, Andrew D White, Andrew J Keefe & Shaoyi Jiang,(2012) "Sequence,  structure, and function of peptide self-assembled monolayers", Journal of the American Chemical Society, Vol.134, Issue 13, pp 6000-6005.
[4]     Andrew David White, Ann Kasia Nowinski, Wenjun Huang, Andrew Keefe, Fang Sun & Shaoyi Jiang, (2012) "Decoding nonspecific interactions from nature", Chemical Science, Issue 12, pp3488-3494.
[5]     Jinjun Zhang,Bonsung Koo,Nithya Subramanian,Yingtao Liu & Aditi Chattopadhyay, (2015) "An optimized cross-linked network model to simulate the linear elastic material response of a smart polymer", Journal of Intelligent Material Systems and Structures.
[6]     Jinjun Zhang, Bonsung Koo, Yingtao Liu, Jin Zou, Aditi Chattopadhyay & Lenore Dai, (2015) "A novel statistical spring-bead based network model for self-sensing smart polymer materials." Smart Materials and Structures, Volume 24, Issue8.
[7]     Jinjun Zhang, Kuang Liu, Chuntao Luo & Aditi Chattopadhyay, (2013) "Crack initiation and fatigue life prediction on aluminum lug joints using statistical volume element‐based multiscale modeling", Journal of Intelligent Material Systems and Structures volume 24, Issue 17 pp 2097-2109.
[8]     Jinjun Zhang, J. Johnston & Aditi Chattopadhyay, (2014)"Physics‐based multiscale damage criterion for fatigue crack prediction in aluminium alloy", Fatigue & Fracture of Engineering Materials & Structures, volume 37, issue 2, pp119-131.
[9]     JiaoJiao Wang, AB Phillion & GuiMin Lu, (2015)"Development of a visco-plastic constitutive modeling for thixoforming of AA6061 in semi-solid state", Journal of Alloys and Compounds, volume 609, pp 290-295
[10]    JiaoJiao Wang, D Brabazon, AB Phillion & GuiMin Lu, (2015) "An innovative two-stage reheating process for wrought aluminum alloy during thixoforming", Metallurgical and Materials Transactions A, Volume 46, Issue 9, pp 4191-4201

[11]     Jiaojiao Wang, Shuzhen Shang, Guimin Lu & Jianguo Yu, (2013) "Viscosity estimation of semi-solid alloys based on thermal simulation compression tests", International Journal of Materials Research,Volume 104, Issue 3, , pp 255-259

[12]     Jiao Jiao Wanga, Zhong Min Zhang, AB Phillion, Shu Zhen Shang & Gui Min Lub, (2016)"Alloy development and reheating process exploration of Al–Si casting alloys with globular grains for thixoforming",J. Mater. Res,

[13]     Gang Wang, Jingxian Wu, Guoqing Zhou & Geoffrey Ye Li, (2013) "Collision-tolerant media access control for asynchronous users over frequency-selective channels," IEEE Transactions on Wireless Communications, vol. 12, no. 10, pp 5162-5171.

[14]     Gang Wang, Jingxian Wu & Yahong Rosa Zheng, (2014)"Optimum energy and spectral efficient transmissions for delay-constrained hybrid ARQ systems," IEEE Transactions on Vehicular Technology.

[15]     Cheng Li & Paul Ampadu, (2015) "Energy-efficient NoC with variable channel width", IEEE 58th International Midwest Symposium on Circuits and Systems (MWSCAS).

[16]     Cheng Li & Paul Ampadu, (2015) "A compact low-power eDRAM-based NoC buffer", IEEE/ACM International Symposium on Low Power Electronics and Design (ISLPED), pp 116-121

[17]     Liangliang Zhang, Yuzheng Guo, Vinayak Vishwanath Hassan, Kechao Tang, Majeed A. Foad, Joseph C. Woicik, Piero A. Pianetta, John Robertson & Paul C McIntyre, (2016) "Interface Engineering for Atomic Layer Deposited Alumina Gate Dielectric on SiGe Substrates", ACS Applied Materials & Interfaces, Volume 8,Issue 29, pp 19110-19118.

[18]     Liangliang Zhang, Huanglong Li, Yuzheng Guo, Kechao Tang & Paul C. McIntyre, (2015) "Selective Passivation of GeO2/Ge Interface Defects in Atomic Layer Deposited High-k MOS Structures", in ACS Applied Materials & Interfaces, Volume 7, Issue 37, pp 20499–20506.

[19]     Pei Luo, Yunsi Fei, Xin Fang, A Adam Ding, David R Kaeli & Miriam Leeser, (2015) "Side-Channel Analysis of MAC-Keccak Hardware Implementations", Proceedings of the Fourth Workshop on Hardware and Architectural Support for Security and Privacy.

[20]     Tushar Swamy, Neel Shah, Pei Luo, Yunsi Fei, David Kaeli, (2014) "Scalable and efficient implementation of correlation power analysis using graphics processing units (GPUs)", Proceedings of the Third Workshop on Hardware and Architectural Support for Security and Privacy.

[21]     Shaohui Wang & Bing Wei, (2015) "Multiplicative Zagreb indices of k-trees", Discrete Applied Mathematics 180, pp 168-175.

[22]     Chunxiang Wang, Shaohui Wang & Bing Wei, (2016) "Cacti with Extremal PI Index", Transactions on Combinatorics, Volume 5, Issue 4.

[23]     Jiaze He & Fuh-Gwo Yuan, (2015) "Damage Identification for Composite Structures using a Cross-Correlation Reverse-Time Migration Technique", Structural Health Monitoring, 14(6): pp 558-570.

[24]     Jiaze He & Fuh-Gwo Yuan, (2016) "Lamb wave-based subwavelength damage imaging using the DORT-MUSIC technique in metallic plates." Structural Health Monitoring, 15(1) pp 65–80.

[25]     Vinay Narayanunni, Heng Gu & Choongho Yu, (2011)"Monte Carlo Simulation for Investigating Influence of Junction and Nanofiber Properties on Electrical Conductivity of Segregated-network Nanocomposites"; Acta Materialia, Volume 59, Issue 11, pp 4548–4555.

[26]     Yihua Zhou, Walter Hu, Bei Peng & Yaling Liu, (2014)"Biomarker Binding on an Antibody Functionalized Biosensor Surface: The Influence of Surface Properties, Electric Field, and Coating Density", The Journal of Physical Chemistry C, pp14586-14594.

[27]     Ru-Ze Liang, Lihui Shi, Haoxiang Wang, Jiandong Meng, Jim Jing-Yan Wang, Qingquan Sun & Yi Gu, (2016)"Optimizing Top Precision Performance Measure of Content-Based Image Retrieval by Learning Similarity Function", 23st International Conference on Pattern Recognition.

[28]     Ru-Ze Liang, Wei Xie, Weizhi Li, Hongqi Wang, Jim Jing-Yan Wang, Lisa Taylor, (201) "A novel transfer learning method based on common space mapping and weighted domain matching", IEEE 28th International Conference on Tools with Artificial Intelligence (ICTAI)

[29]     Wei Cai & Leslie Lauren Gouveia, "Modeling and simulation of Maximum power point tracker in Ptolemy" , Journal of Clean Energy Technologies, Vol. 1, No. 1, 2013 , PP 6-9.

[30]     Wei Cai, Jeremy. Chan & David Garmire, "3-Axes MEMS Hall-Effect Sensor," presented by the 2011 IEEE Sensors Applications Symposium, pp141-144.

[31]   Wei Cai, Xuelin Cui & Xiangrong Zhou, "Optimization of a GPU Implementation of Multi-dimensional RF Pulse Design Algorithm," International Conference on Bioinformatics and Biomedical Engineering 2011
[32]   Wei Cai & Frank Shi, (2016) "2.4 GHz Heterodyne Receiver for Healthcare Application", IJPPS, vol 6,pp 1-7.
[33]   Kunal Datta & Hossein Hashemi, ( 2014) "Performance Limits, Design and Implementation of mm-Wave SiGe HBTClass-E and Stacked Class-E Power Amplifiers," JSCC, vol 49, pp 2150-2171.
[34]   Wei Cai & Frank Shi, (2016) "High Performance SOI RF Switch for Healthcare Application", 2016，International Journal of Enhanced Research in Science, Technology & Engineering，Volume 5，Issue 10, pp 23-28.
[35]   P .Manikandan & Ribu Mathew, (2012) "Design of CMOS Class-E Power Amplifier for WLAN and Bluetooth",  ICDCS,  pp 81-88,.
[36]   Wei cai, liang huang & Wujie Wen, "2.4GHZ Class AB Power  Amplifier for Healthcare Application" International Journal of Biomedical Engineering and Science (IJBES), Vol. 3, No. 2, April 2016

**Authors**

Wei Cai is a graduate student at the University of California, Irvine, CA. She received her Masters degree from Dept. of Electrical Engineering, University of Hawaii at Manoa and Bachelor degree from Zhejiang University, China. Her research interests include device physics simulation, analog/ RF circuit design.

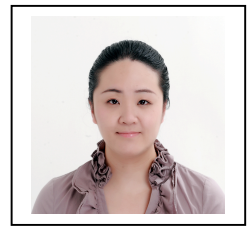

Jian Xu is an assistant professor at the Division of Electrical and Computer Engineering in the School of Electrical Engineering and Computer Science of Louisiana State University. He got his Ph. D. in Electrical Engineering at Yale University. He received a B. S. and a M. S. degree in Physics at Nanjing University, China. His research interest is in the Bioelectronics, Nanoelectronics and Nanomedicine, and Biomedical instrumentation for image-guided cancer surgery.

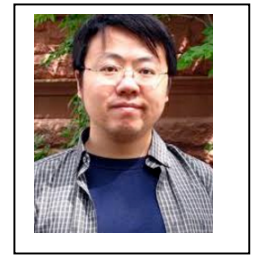

Liang Huang is an associate Professor,   Electronics College of Zhejiang Gongshang University. He got phd from Zhejiang University china, and finished his postdoc at Polytechnic of  Turin, Italy, and Hanyang University, Seoul, Korea. His research is mainly focus on Research on: Intelligent Control; Electrical Robotics.

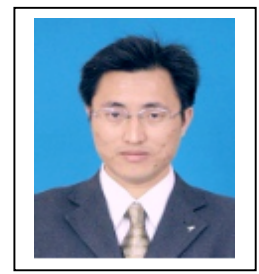